\documentclass[journal]{IEEEtran}
\usepackage[T1]{fontenc}
\usepackage[latin9]{inputenc}
\usepackage{array}
\usepackage{mathtools}
\usepackage{algorithm2e}
\usepackage{amsmath,amssymb}
\usepackage{graphicx}
\usepackage{xcolor,color,colortbl}
\definecolor{LightCyan}{rgb}{0.9,1,1}
\definecolor{LightGreen}{rgb}{0.9,1,0.9}
\definecolor{LightYellow}{rgb}{1,1,0.9}
\definecolor{Gray}{gray}{0.9}
\usepackage[unicode=true,
 bookmarks=true,bookmarksnumbered=true,bookmarksopen=true,bookmarksopenlevel=1,
 breaklinks=false,pdfborder={0 0 1},backref=false,colorlinks=false]
 {hyperref}
\hypersetup{pdftitle={Your Title},
 pdfauthor={Your Name},
 pdfborderstyle=,pdfpagelayout=OneColumn,pdfnewwindow=true,pdfstartview=XYZ,plainpages=false}

\makeatletter

\providecommand{\tabularnewline}{\\}

\makeatother

\begin{document}
\title{Systematic Design and Optimization of Quantum Circuits for Stabilizer Codes}
\author{Arijit Mondal and Keshab K. Parhi, {\em Fellow, IEEE}\\
 Email: \{monda109, parhi\}@umn.edu\\
 Department of Electrical and Computer Engineering, University of
Minnesota}
\maketitle
\begin{abstract}
Quantum computing is an emerging technology that has the potential to achieve exponential speedups over their classical counterparts. To achieve quantum advantage, quantum principles are being applied to
fields such as communications, information processing, and artificial intelligence. However, quantum computers face a fundamental issue since quantum bits are extremely noisy and prone to decoherence. Keeping qubits error free is one of the most important steps towards reliable quantum
computing. Different stabilizer codes for quantum error correction have been proposed in past decades and several methods have been proposed to import classical error correcting codes to the quantum domain. However, formal approaches towards the design and optimization of circuits for these quantum encoders and decoders have so far not been proposed. In this paper, we propose a formal algorithm for systematic construction of encoding circuits for general stabilizer codes. This algorithm is used to design encoding and decoding circuits for an eight-qubit code. Next, we propose a systematic method for the optimization of the encoder circuit thus designed. Using the proposed method, we optimize the encoding circuit in terms of the number of 2-qubit gates used. The proposed optimized eight-qubit encoder uses 18 CNOT gates and 4 Hadamard gates, as compared to 14 single qubit gates, 33 2-qubit gates, and 6 CCNOT gates in a prior work. The encoder and decoder circuits are verified using IBM Qiskit. We also present optimized encoder circuits for Steane code and a 13-qubit code in terms of the number of gates used.
\end{abstract}


\section{Introduction}

Quantum computing is a rapidly-evolving technology which exploits
the fundamentals of quantum mechanics towards solving tasks which
are too complex for current classical computers.
In 1994, a quantum algorithm to find the prime
factors of an integer in {\em polynomial} time was proposed by Shor \cite{shor1994algorithms} .
In 1996, a quantum algorithm to search a particular
element in an unsorted database with a high probability and significantly
higher efficiency than any known classical algorithm was presented by Grover \cite{grover1996fast}.
The realization of these powerful algorithms requires massive quantum
computers with the capability of processing a large number of qubits. 

The phenomenon through which quantum mechanical systems attain interference
among each other is known as quantum coherence. Quantum coherence
is essential to perform quantum computations on quantum information.
However, quantum systems are inherently susceptible to noise and decoherence.
Maintaining coherence and mitigating noise becomes increasingly challenging 
with the increase in the number of qubits in a quantum computer. 
Thus, quantum error correcting codes (ECCs)
become essential for reliable quantum computing systems. There were various
challenges in the process of designing a quantum ECC framework. It is
well known that measurement destroys superposition in any quantum
system. Also, since the quantum errors are continuous in nature, the
design of an ECC for quantum systems was difficult. Furthermore, the no-go theorems in the quantum domain make it difficult to
design an ECC system analogous to classical domain \cite{nielsen,wootters,dieks,kumar,barnum}.

Quantum ECCs were
believed to be impossible till 1995, when Shor demonstrated a 9-qubit
ECC which was capable of correcting a single qubit error for the first
time \cite{shor}. In 1996, Gottesman proposed a stabilizer framework
which was widely used for construction of quantum ECCs from classical
ECCs \cite{gottesman,gottesman-thesis}. Calderbank-Shor-Steane (CSS)
codes were proposed independently by Calderbank-Shor \cite{calderbank}
and Steane \cite{steane}. These codes were used to derive quantum
codes from binary classical linear codes. 
Pre-shared entangled qubits were used to
construct stabilizer codes over non-Abelian groups in \cite{brun2006correcting}.
These entanglement-assisted
(EA) stabilizer codes contain qubits over the extended operators which
are assumed to be at the receiver end throughout, and entangled with
the transmitted set of qubits. It was later shown that EA stabilizer
codes increase the error correcting capability of quantum ECCs \cite{lai2013entanglement}.

An encoding
procedure for EA stabilizer codes were proposed in \cite{wilde2008quantum}. Quantum low-density parity-check (LDPC) codes were constructed from classical quasi-cyclic binary LDPC codes by the authors in \cite{hsieh2009entanglement}. Quantum analogs of Reed Solomon (RS) codes were constructed from self-orthogonal classical RS codes in \cite{grassl1999quantum,aly2008asymmetric,la2012asymmetric}.
Purely quantum polar codes based on recursive channel combining and splitting construction were studied in \cite{dupuis2019purely}. EA stabilizer codes were
extended to qudit systems in \cite{nadkarni2021entanglement}. Recently, a universal decoding scheme was conceived
for quantum stabilizer codes (QSCs) by adapting `guessing random additive noise decoding' (GRAND) philosophy from classical domain codes \cite{chandra2023universal}. However, it becomes necessary to design and simulate actual encoder and decoder
circuits for these quantum ECCs, so that reliable quantum computing
systems can be built. The CSS framework is particularly interesting
due to its simplicity as it is useful for importing classical codes to quantum domain if they satisfy certain properties \cite{calderbank,steane}.

Though a lot of papers have been published in the field of quantum ECCs, design of circuits to implement those quantum ECCs have not been explored to that extent. The primary contribution of this paper is a systematic method for the design and optimization of encoder-decoder circuits for stabilizer codes \footnote{EA stabilizer codes require pre-shared entangled qubits to be present between the encoder and decoder. Due to this constraint, we consider general stabilizer codes instead of EA stabilizer codes in this work.}, along with the simulation of the proposed circuits using IBM Qiskit \cite{qiskit}. A systematic method for the construction of an encoder for stabilizer codes was demonstrated for a five qubit code, which uses five physical qubits to encode a logical qubit, in \cite{gottesman-thesis}. We review this construction and propose a formal algorithm for the construction of encoder circuits for stabilizer codes. Next, we apply it to the eight-qubit code $[[8,3,3]]$ which encodes three logical qubits using eight physical qubits and is more efficient in terms of code rate than the 5-qubit code. 

CNOT gates play a vital role as building blocks for quantum circuits. It can be shown that any arbitrary unitary transformation on a $n$-qubit system can be performed as a combination of CNOT gates and other single qubit unitary gates \cite{nielsen}, thus forming an universal set of gates for quantum circuits. The circuits designed in \cite{gottesman-thesis} contains $3$ types of $2$-qubit gates, the CNOT, CZ, and the CY gates. However, for practical quantum circuits, it is more convenient to have a single type of 2-qubit gates between any arbitrary pair of qubits. 
Similar to equivalence rules in classical digital circuits, several equivalencies also exist for quantum circuits. These were used to study quantum circuits in \cite{zhou2000methodology,mermin2001classical,mermin2002deconstructing,maslov2008quantum}. The authors in \cite{garcia} derived additional rules and compiled those into a set of equivalence rules. In the current experimental quantum circuits, the error rates of the CNOT gates are one of the major causes of circuit reliability~\cite{linke2017experimental,wright2019benchmarking,bataille2022quantum}.Thus, it becomes extremely important to optimize the quantum circuits in terms of the number of CNOT gates. There exists a one-to-one correspondence between $n$-qubit systems consisting of CNOT gates and $n\times n$ non-singular matrices with coefficients in $\mathbb{F}_2$  as proved in \cite{alber2001quantum,patel2008optimal}. The authors in \cite{bataille2022quantum} described this correspondence in a formal way in terms of group isomorphism and group representation. They used this approach to propose an optimization algorithm for a system consisting of a few qubits, with CNOT gates interacting between them.

The contributions of this paper are four-fold. First, we revisit the systematic method
for construction of encoder and decoder circuits for stabilizer codes. We identify and analyze the key concepts for the construction of an encoder for stabilizer codes, demonstrated in \cite{gottesman-thesis} through a five-qubit code. The concepts are then used to formulate an algorithm for the construction of an encoder circuit for a general stabilizer code. Second, the encoder circuit for the eight-qubit encoder designed in this paper is more efficient compared to a prior design \cite{dong2013efficient} in terms of number of gates used. The construction in \cite{dong2013efficient} requires two times more gates than the circuit designed in this paper. Third, we optimize the encoder circuit we designed in terms of number of gates used. We use only CNOT and $H$ gates for the circuit design. To the best of our knowledge, such an optimized circuit for stabilizer codes has not been explored before. For the decoder designs, we use a syndrome measurement circuit, and
depending on the measured syndromes, we apply the appropriate error correction using suitable Pauli gates. Fourth, we design encoder circuits for Steane code \cite{steane1996multiple} and a 13-qubit code and optimize those circuits in terms of number of gates used. Finally, we verify all the circuits we designed using IBM Qiskit \cite{qiskit}.

The rest of the paper is organized as follows. In Section II, we give
a brief theory of the stabilizer framework which is essential for
the design of the quantum ECCs. We also provide the theoretical background
of the CSS framework. We then propose an algorithm for systematic construction of an encoder circuit for a general stabilizer code. In Section III, we provide detailed
procedures for the construction of the encoder and decoder circuits
for the eight-qubit code. Section IV deals with the optimization of the circuits for the 8-qubit code we designed in Section III. In Section V, we discuss the design of encoder circuit for Steane code and a 13-qubit code, and also present optimized circuits for those encoders. We
discuss the results and comparisons in Section VI. Section VII concludes the paper.

\section{Theoretical background}

Shor's 9-qubit code \cite{shor} was the first ever quantum ECC capable
of correcting a single qubit error. Gottesman \cite{gottesman-thesis}
proposed a general methodology to construct quantum
ECCs. This method is known as the stabilizer construction and the
codes thus generated are known as stabilizer codes. Before going into
the details of stabilizer codes, we briefly describe a quantum
state and explain how the states evolve in a quantum circuit. 

In two-level quantum systems, the two-dimensional unit of quantum
information is called a quantum bit (qubit). The state of a qubit
is represented by $|\psi\rangle=a|0\rangle+b|1\rangle,$ where $a,b\in \mathbb{C}$
and $|a|^{2}+|b|^{2}=1$. $|0\rangle$ and $|1\rangle$ are basis
states of the state space. The evolution of a quantum mechanical system
is fully described by a unitary transformation. State $|\psi_{1}\rangle$
of a quantum system at time $t_{1}$ is related to state $|\psi_{2}\rangle$ at time $t_{2}$ by a unitary
operator $U$ that depends only on the time instances $t_{1}$ and
$t_{2}$, i.e., $|\psi_{2}\rangle=U|\psi_{1}\rangle$. The unitary
operators or matrices which act on the qubit belong to $\mathbb{C}^{2\times2}$.
We have a Pauli group which represents the unitary matrices given
by

\begin{equation} \label{eq:Pauli}
\Pi=\{\pm I_{2},\pm iI_{2},\pm X,\pm iX,\pm Y,\pm iY,\pm Z,\pm iZ\}
\end{equation}

where $I_{2}=\left[\begin{array}{cc}
1 & 0\\
0 & 1
\end{array}\right],\,X=\left[\begin{array}{cc}
0 & 1\\
1 & 0
\end{array}\right],\,Y=\left[\begin{array}{cc}
0 & -i\\
i & 0
\end{array}\right],\,Z=\left[\begin{array}{cc}
1 & 0\\
0 & -1
\end{array}\right]$.

A quantum circuit consists of an initial set of qubits as inputs which
evolve through time to a final state, comprising of
the outputs of the quantum circuit. Quantum states evolve through
unitary operations which are represented by quantum gates. Quantum
gates can be single qubit gates which act on a single qubit, or they
can be multiple qubit gates which act on multi-qubit states to produce
a new multi-qubit state. The single qubit gates include the bit flip gate
$X$, phase flip gate $Z$, Hadamard gate $H$, $Y$ gate, and the
phase gate $S$. The unitary operations related to the single qubit
gates are as follows:

\begin{align}
X & =\left[\begin{array}{cc}
0 & 1\\
1 & 0
\end{array}\right],Z=\left[\begin{array}{cc}
1 & 0\\
0 & -1
\end{array}\right],H=\frac{1}{\sqrt{2}}\left[\begin{array}{cc}
1 & 1\\
1 & -1
\end{array}\right],\\
Y & =\left[\begin{array}{cc}
0 & -i\\
i & 0
\end{array}\right],S=\left[\begin{array}{cc}
1 & 0\\
0 & i
\end{array}\right]\nonumber 
\end{align}

The multi-qubit gates include a controlled-$X$ (CNOT), controlled-$Z$ (CZ),
and controlled-$Y$ (CY) gates. They act on 2-qubit states and are given by the
following unitary transformations:

\begin{equation}
CNOT=\left[\begin{array}{cccc}
1 & 0 & 0 & 0\\
0 & 1 & 0 & 0\\
0 & 0 & 0 & 1\\
0 & 0 & 1 & 0
\end{array}\right],CZ=\left[\begin{array}{cccc}
1 & 0 & 0 & 0\\
0 & 1 & 0 & 0\\
0 & 0 & 1 & 0\\
0 & 0 & 0 & -1
\end{array}\right],
\end{equation}

\begin{equation}
CY=\left[\begin{array}{cccc}
1 & 0 & 0 & 0\\
0 & 1 & 0 & 0\\
0 & 0 & 0 & -i\\
0 & 0 & i & 0
\end{array}\right]
\end{equation}

\begin{figure}
\begin{centering}
\includegraphics[scale=0.5]{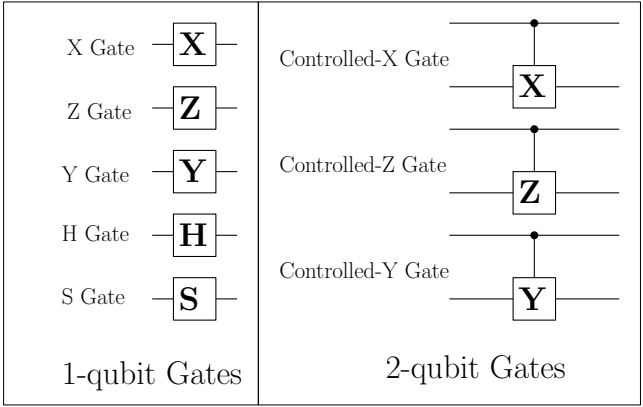}
\par\end{centering}
\caption{Symbolic representations of various 1-qubit and 2-qubit gates. \label{fig:gates}}
\end{figure}

Symbolic representations of various 1-qubit and 2-qubit gates are shown in Fig. \ref{fig:gates}. For further understanding on quantum circuits and introduction to quantum ECCs, interested readers are referred to \cite{steane2006tutorial,roffe2019quantum}. With the above background, we are ready to describe the stabilizer
formalism for quantum ECCs.

\subsection{Quantum code and Stabilizer formalism}

An $[[n,k]]$ quantum code can be used
for quantum error correction, where $k$ logical qubits are encoded using
$n$ physical qubits, leading to a code rate of $k/n$ analogous to classical
error correction. It has $2^k$ basis codewords, and any linear combination of the basis codewords are also valid codewords. Let the space of valid codewords be denoted by $T$. If we consider the tensor product of Pauli operators (with possible overall factors of $\pm 1$ or $\pm i$) in equation \ref{eq:Pauli}, it forms a group $G$ under multiplication. The stabilizer $S$ is
an Abelian subgroup of $G$, such that the code space $T$ is the space of vectors fixed by $S$ \cite{gottesman,gottesman-thesis}. Stabilizer generators are a set of independent set of $n-k$ elements from the stabilizer group, in the sense that none of them is a product of any two other generators.

We know that the operators in the Pauli group act on single qubit
states which are represented by $2$-element vectors. The operators in
$\Pi$ have eigen values $\pm1$, and either commute or anti-commute
with other elements in the group. The set $\Pi^{n}$ is given by the
$n$-fold tensor products of elements from the Pauli group $\Pi$
as shown below,

\begin{align}
\Pi^{n}= & \{e^{i\phi}A_{1}\otimes A_{2}\otimes\cdots\otimes A_{n}\nonumber \\
 & :\forall j\in\{1,2,\cdots,n\}A_{j}\in\Pi,\phi\in\{0,\pi/2,\pi,3\pi/2\}\}
\end{align}

The stabilizer is a group with elements $M$ such that $M|\psi\rangle=|\psi\rangle$.
The stabilizer is Abelian, i.e., every pair of elements in the stabilizer
group commute. This can be verified from the following observation.
If $M|\psi\rangle=|\psi\rangle$ and $N|\psi\rangle=|\psi\rangle$,
then $MN|\psi\rangle-NM|\psi\rangle=(MN-NM)|\psi\rangle=0$. Thus,
$MN-NM=0$ or $MN=NM$, showing that every pair of elements in the
stabilizer group commute.

Given an Abelian subgroup $S$ of $n$-fold Pauli operators, code space is defined
as

\begin{equation}
T(S)=\{|\psi\rangle,s.t.\,M|\psi\rangle=|\psi\rangle,\forall M\in S\}
\end{equation}

Suppose $M\in S$ and Pauli operator $E$ anti-commutes with $M$.
Then, $M(E|\psi\rangle)=-EM|\psi\rangle=-E|\psi\rangle$. Thus, $E|\psi\rangle$
has eigenvalue $-1$ for $M$. Conversely, if Pauli operator $E$
commutes with $M$, $M(E|\psi\rangle)=EM|\psi\rangle=E|\psi\rangle$;
thus $E|\psi\rangle$ has eigenvalue $+1$ for $M$. Therefore, eigenvalue
of an operator $M$ from a stabilizer group detects errors which anti-commute
with $M$.

\subsection{Binary vector space representation for stabilizers} \label{sec-2-b}

The stabilizers can be written as binary vector spaces, which can be useful to bring connections with classical error correction theory \cite{gottesman-thesis}. For this, the stabilizers are written as a pair of $(n-k)\times n$ matrices. The rows correspond to the stabilizers and the columns correspond to the qubits. The first matrix has a $1$ wherever there is a $X$ or $Y$ in the corresponding stabilizer, and $0$  everywhere else. We will refer to this as the $X$-portion of the matrix. The second matrix has a $1$ wherever there is a $Z$ or $Y$ in the corresponding stabilizer and $0$ everywhere else. We will refer to this as the $Z$-portion of the matrix. It is often more convenient to write the two matrices as a single $(n-k)\times 2n$ matrix with a vertical line separating the two.

\subsection{CSS framework}

The CSS framework \cite{calderbank,steane} is a method to construct quantum ECCs from
their classical counterparts. We can combine classical codes with parity check matrices $P_1$ and $P_2$ into a quantum code if the rows of $P_1$ and $P_2$  are orthogonal using the binary dot product. This implies that dual-containing codes can be imported to the quantum domain. Given two classical codes $C_{1}[n,k_{1},d_{1}]$
and $C_{2}[n,k_{2},d_{2}]$ which satisfy the dual containing criterion
$C_{1}^{\perp}\subset C_{2}$, CSS framework can be used to construct
quantum codes from such codes.

The CSS codes form a class of stabilizer codes. From the classical
theory of error correction, let $H_{1}$ and $H_{2}$ be the check
matrices of the codes $C_{1}$ and $C_{2}$. Since $C_{1}^{\perp}\subset C_{2}$,
codewords of $C_{2}$ are basically the elements of $C_{1}^{\perp}$.
Hence, we have, $H_{2}H_{1}^{T}=0$. The check matrix of a CSS code
is given by:

\begin{equation}
H_{C_{1}C_{2}}=\left[\begin{array}{ccc}
\begin{array}{c}
H_{1}\\
0
\end{array} & \Bigg| & \begin{array}{c}
0\\
H_{2}
\end{array}\end{array}\right]
\end{equation}

\subsection{Systematic procedure for encoder design for a stabilizer code}\label{sec-3-a} \label{sec:2-d}

A systematic method for the design of an encoder for a stabilizer code was presented in \cite{gottesman-thesis}. We make slight modifications to the method, and propose a complete procedure for the design of an encoder circuit for a stabilizer code can be summarized as follows:

\textbf{Step 1}: The stabilizers are written in a matrix form using binary vector space formalism as mentioned in Section \ref{sec-2-b}. Let the parity chek matrix thus obtained be $H_{q}$.

\textbf{Step 2}: Our aim is to bring $H_{q}$ to the standard form $H_{s}$ below:
\begin{equation} \label{eq:std-hs}
H_{s}=\left[\begin{array}{ccc}
\begin{array}{ccc}
I_{1} & A_{1} & A_{2}\\
0 & 0 & 0
\end{array} & \Bigg| & \begin{array}{ccc}
B & C_{1} & C_{2}\\
D & I_{2} & E
\end{array}\end{array}\right]
\end{equation}

where, $I_{1}$ and $B$ are $r\times r$ matrices. `$r$' is the rank of the $X$ portion of $H_{s}$. $A_{1}$ and $C_{1}$
are $r\times(n-k-r)$ matrices. $A_{2}$ and $C_{2}$ are $r\times k$
matrices. $D$ is a $(n-k-r)\times r$ matrix. $I_{2}$ is a $(n-k-r)\times(n-k-r)$
matrix. $E$ is a $(n-k-r)\times k$ matrix. $I_{1}$ and $I_{2}$
are identity matrices. 

$H_{q}$ is converted to standard form $H_{s}$ using Gaussian elimination \cite{gottesman-thesis}. The logical operators $\overline{X}$ and $\overline{Z}$ can be written as 
\begin{equation}
\overline{X}=\left[\begin{array}{ccc}
\begin{array}{ccc}
0 & U_{2} & U_{3}\end{array} & | & \begin{array}{ccc}
V_{1} & 0 & 0\end{array}\end{array}\right]
\end{equation}

\begin{equation}
\overline{Z}=\left[\begin{array}{ccc}
\begin{array}{ccc}
0 & 0 & 0\end{array} & | & \begin{array}{ccc}
V'_{1} & 0 & V'_{3}\end{array}\end{array}\right]
\end{equation}

where $U_{2}=E^T$, $U_{3}=I_{k\times k}$, $V_{1}=E^{T}C_1^{T}+C_2^{T}$, $V'_{1}=A_{2}^{T}$, and $V'_{3}=I_{k\times k}$.

Given the parity check matrix in standard form $H_{s}$ and $\overline{X}$, the encoding operation for a stabilizer code can be written as,

\begin{align}
|c_{1}c_{2}\cdots c_{k}\rangle= & \overline{X}_{1}^{c_{1}}\overline{X}_{2}^{c_{2}}\cdots \overline{X}_{k}^{c_{k}}\left(\sum_{M\in S}M\right)|00\cdots 0\rangle\\
= & \overline{X}_{1}^{c_{1}}\overline{X}_{2}^{c_{2}}\cdots \overline{X}_{k}^{c_{k}}(I+M_{1})(I+M_{2})\cdots\nonumber \\
 & (I+M_{n-k})|00\cdots 0\rangle.
\end{align}

There are a total of $n$ qubits. Place qubits initialized to $|0\rangle$ at qubit positions $i=1$ to $i=n-k$. Place the qubits to be encoded at positions $i=n-k+1$ to $i=n$.

We observe the following from $H_{s}$ and $\overline{X}$:

\begin{itemize}

\item We know that a particular logical operator $\overline{X}_i$ is applied only if the qubit at $i^{\mathrm{th}}$ position is $|1\rangle$. Thus, applying $\overline{X}_i$ controlled at $i^{\mathrm{th}}$ qubit encodes $\overline{X}_i$.

\item The $\overline{X}$ operators consist of products of only $Z$s for the first $r$ qubits. For the rest of the qubits, $\overline{X}$ consists of products of $X$s only. We know that $Z$ acts trivially on $|0\rangle$. Since the first $r$ qubits are initialized to $|0\rangle$, we can ignore all the $Z$s in $\overline{X}$. 

\item The first $r$ generators in $H_{s}$ apply only a single bit flip to the first $r$ qubits. This implies that when $I+M_{i}$ is applied, the resulting state would be a sum of $|0\rangle$ and  $|1\rangle$ for the $i^{\mathrm{th}}$ qubit. This corresponds to applying $H$ gates to the first $r$ qubits, which puts each of the $r$ qubits in the state $\frac{1}{\sqrt{2}}(|0\rangle+|1\rangle)$.

\item If we apply $M_{i}$ conditioned on qubit $i$, it implies the application of $I+M_{i}$. The reason is as follows. When the control qubit $i$ is $|1\rangle$, $M_i$ needs to be applied to the combined qubit state. Since the qubit $i$ suffers a bit flip $X$ only by the stabilizer $M_{i}$, it is already in flipped state when it is $|1\rangle$. Thus, only the rest of the operators in $M_{i}$ need to be applied. However, there would be an issue if $H_{s_{(i,i+n)}}$ is not $0$, i.e., there is a $Y$ instead of $X$. In that case, adding an $S$ gate after the $H$ gate resolves the issue.

\end{itemize}

\textbf{Step 3}: The observations in Step 2 can be used to devise an algorithm as shown in Algorithm \ref{alg:algo-1} to design the encoding circuit.

\RestyleAlgo{ruled}
\begin{algorithm}
\caption{Algorithm to generate encoding circuit from $H_{s}$ and $\overline{X}$ ($n$ = number of physical qubits, $k$ = number of logical qubits, $r$ = rank of $X$-portion of $H_{s}$).}\label{alg:algo-1}
\KwData{$H_{s}$, $\overline{X}$}
\KwResult{Encoding circuit}
 \For{$i=1$ \KwTo $k$}{ 
    \If{$\overline{X}_{i,i+n-k}==1$}{
        Place controlled dot at qubit $i+n-k$\
        }
    \For{$j=1$ \KwTo $n$}{
        \If{$i+n-k\neq j$}{
            \If{$\overline{X}_{i,j}==1$}{
                Place $X$ gate at qubit $j$ controlled at qubit $i+n-k$
            }
        }
    }
 }
 \For{$i=1$ \KwTo $r$}{ 
    \eIf{$H_{s_{(i,i+n)}}==0$}{
        Place $H$ gate followed by controlled dot at qubit $i$\
        }
        {
        Place $H$ gate followed by $S$ gate followed by controlled dot at qubit $i$\
        }
    \For{$j=1$ \KwTo $n$}{
        \If{$i\neq j$}
            {\If{$H_{s_{(i,j)}}==1$ \&\& $H_{i,j+n}==0$}{
                Place $X$ gate on qubit $j$ with control at qubit $i$\
                }
            \If{$H_{s_{(i,j)}}==0$ \&\&  $H_{i,j+n}==1$}{
                Place $Z$ gate on qubit $j$ with control at qubit $i$\
                }
            \If{$H_{s_{(i,j)}}==1$ \&\&  $H_{i,j+n}==1$}{
                Place $Y$ gate on qubit $j$ with control at qubit $i$\
                }
            }
        }
    }
    
\end{algorithm}

Minor modifications can be done to Algorithm \ref{alg:algo-1} to design an encoder circuit that uses CNOT and CZ gates only as multiple qubit gates. For this, instead of CY gates, controled-XZ gates are used. Also, the $S$ gate is replaced with a $Z$ gate. The rest of the algorithm remains the same.

\section{Encoder and decoder circuit design for eight-qubit code {[}{[}8,3,3{]}{]}}

An eight-qubit code encodes 3 logical qubits using 8 physical qubits
and can correct a single qubit error. The eight-qubit code is represented
by the following stabilizer generators \cite{gottesman-thesis}:

\begin{center}
\begin{tabular}{c|cccccccc}
$M_{1}$ & $X$ & $X$ & $X$ & $X$ & $X$ & $X$ & $X$ & $X$\tabularnewline
$M_{2}$ & $Z$ & $Z$ & $Z$ & $Z$ & $Z$ & $Z$ & $Z$ & $Z$\tabularnewline
$M_{3}$ & $I$ & $X$ & $I$ & $X$ & $Y$ & $Z$ & $Y$ & $Z$\tabularnewline
$M_{4}$ & $I$ & $X$ & $Z$ & $Y$ & $I$ & $X$ & $Z$ & $Y$\tabularnewline
$M_{5}$ & $I$ & $Y$ & $X$ & $Z$ & $X$ & $Z$ & $I$ & $Y$\tabularnewline
\end{tabular}. 
\par\end{center}

\subsection{Encoder design for the eight-qubit code}

The stabilizers for the eight-qubit code can be written in binary vector space formalism, following the process in Section \ref{sec-2-b}. This corresponds to Step 1 in Section \ref{sec-3-a}.

\begin{equation}
H_{q}=\left[\begin{array}{ccc}
\begin{array}{c}
11111111\\
00000000\\
01011010\\
01010101\\
01101001
\end{array} & \Bigg| & \begin{array}{c}
00000000\\
11111111\\
00001111\\
00110011\\
01010101
\end{array}\end{array}\right]
\end{equation}

Step 2 involves converting $H_{q}$ into standard form $H_{s}$ as in equation \ref{eq:std-hs}. First, we push the $2^{\mathrm{nd}}$ row to the $5^{\mathrm{th}}$
position and push the rows below up by one step. We also swap the
$4^{\mathrm{th}}$ column with the $5^{\mathrm{th}}$ column (equivalent
to swapping fourth and fifth qubit position), which would require
the operation of swapping $12^{\mathrm{th}}$ column with $13^{\mathrm{th}}$
column as well. Performing the above operations, we get

\begin{equation}
H_{q}=\left[\begin{array}{ccc}
\begin{array}{c}
11111111\\
01011010\\
01001101\\
01110001\\
00000000
\end{array} & \Bigg| & \begin{array}{c}
00000000\\
00010111\\
00101011\\
01001101\\
11111111
\end{array}\end{array}\right]
\end{equation}

We perform the operation $R_{4}\rightarrow R_{4}+R_{2}$, followed by $R_{2}\rightarrow R_{2}+R_{3}$. Next, we swap $R_{2}$ with $R_{3}$, and $R_{3}$ with $R_{4}$. Finally, performing the operation $R_{1}\rightarrow R_{1}+R_{2}+R_{3}+R_{4}$,
we get the standard form as

\begin{equation}
H_{s}=\left[\begin{array}{ccc}
\begin{array}{c}
10001110\\
01001101\\
00101011\\
00010111\\
00000000
\end{array} & \Bigg| & \begin{array}{c}
01001101\\
00101011\\
01011010\\
00111100\\
11111111
\end{array}\end{array}\right]
\end{equation}

From the standard form $H_{s}$, the stabilizers
which will be used for syndrome measurement are as follows.
\begin{center}
\begin{tabular}{c|cccccccc}
$M_{1}$ & $X$ & $Z$ & $I$ & $I$ & $Y$ & $Y$ & $X$ & $Z$\tabularnewline
$M_{2}$ & $I$ & $X$ & $Z$ & $I$ & $Y$ & $X$ & $Z$ & $Y$\tabularnewline
$M_{3}$ & $I$ & $Z$ & $X$ & $Z$ & $Y$ & $I$ & $Y$ & $X$\tabularnewline
$M_{4}$ & $I$ & $I$ & $Z$ & $Y$ & $Z$ & $Y$ & $X$ & $X$\tabularnewline
$M_{5}$ & $Z$ & $Z$ & $Z$ & $Z$ & $Z$ & $Z$ & $Z$ & $Z$\tabularnewline
\end{tabular}. 
\par\end{center}

From $H_{s}$, we can evaluate the following:

\begin{align}
I_{1} & =\left[\begin{array}{c}
1000\\
0100\\
0010\\
0001
\end{array}\right],A_{1}=\left[\begin{array}{c}
1\\
1\\
1\\
0
\end{array}\right],A_{2}=\left[\begin{array}{c}
110\\
101\\
011\\
111
\end{array}\right],\nonumber \\
B & =\left[\begin{array}{c}
0100\\
0010\\
0101\\
0011
\end{array}\right],C_{1}=\left[\begin{array}{c}
1\\
1\\
1\\
1
\end{array}\right],C_{2}=\left[\begin{array}{c}
101\\
011\\
010\\
100
\end{array}\right],\nonumber \\
D & =\left[1111\right],I_{2}=1,E=\left[111\right]
\end{align}

From the above,

\begin{align}
\overline{X} & =\left[\begin{array}{ccc}
\begin{array}{c}
00001100\\
00001010\\
00001001
\end{array} & \Bigg| & \begin{array}{c}
01100000\\
10010000\\
00110000
\end{array}\end{array}\right]\\
\overline{Z} & =\left[\begin{array}{ccc}
\begin{array}{c}
00000000\\
00000000\\
00000000
\end{array} & \Bigg| & \begin{array}{c}
11010100\\
10110010\\
01110001
\end{array}\end{array}\right]
\end{align}

Thus, the logical $\overline{X}$ and $\overline{Z}$ operators are
\begin{center}
\begin{tabular}{c|cccccccc}
$\overline{X_{1}}$ & $I$ & $Z$ & $Z$ & $I$ & $X$ & $X$ & $I$ & $I$\tabularnewline
$\overline{X_{2}}$ & $Z$ & $I$ & $I$ & $Z$ & $X$ & $I$ & $X$ & $I$\tabularnewline
$\overline{X_{3}}$ & $I$ & $I$ & $Z$ & $Z$ & $X$ & $I$ & $I$ & $X$\tabularnewline
$\overline{Z_{1}}$ & $Z$ & $Z$ & $I$ & $Z$ & $I$ & $Z$ & $I$ & $I$\tabularnewline
$\overline{Z_{2}}$ & $Z$ & $I$ & $Z$ & $Z$ & $I$ & $I$ & $Z$ & $I$\tabularnewline
$\overline{Z_{3}}$ & $I$ & $Z$ & $Z$ & $Z$ & $I$ & $I$ & $I$ & $Z$\tabularnewline
\end{tabular}.
\par\end{center}

Using $H_{s}$ and the $\overline{X}$ operators, the various encoded
states can be evaluated using the following equation

\begin{align}
|c_{1}c_{2}c_{3}\rangle= & \overline{X}_{1}^{c_{1}}\overline{X}_{2}^{c_{2}}\overline{X}_{3}^{c_{3}}\left(\sum_{M\in S}M\right)|00000000\rangle\\
= & \overline{X}_{1}^{c_{1}}\overline{X}_{2}^{c_{2}}\overline{X}_{3}^{c_{3}}(I+M_{1})(I+M_{2})\nonumber \\
 & (I+M_{3})(I+M_{4})|00000000\rangle \label{eq:encoded-states}
\end{align}

Since $M_{5}$ consists of tensor products of $Z$ Pauli operators,
and since $Z$ acts trivially on $|0\rangle$, $I+M_{5}$ which does not
change the state $|00000000\rangle$. Thus, we can ignore $M_{5}$. 

\begin{figure}
\begin{centering}
\includegraphics[scale=0.35]{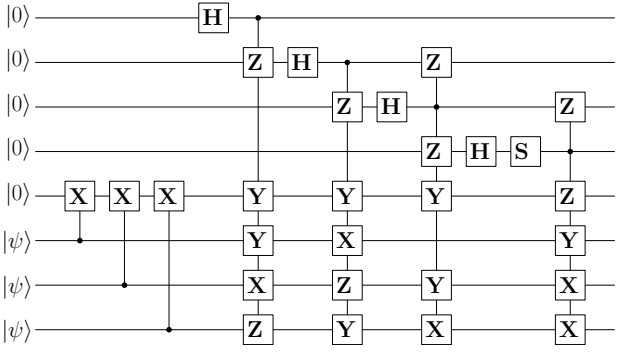}
\par\end{centering}
\caption{Encoder for the eight-qubit code. \label{fig:eight-qubit-encoder}}
\end{figure}

Following the procedure in Section \ref{sec:2-d}, the $3$ qubits to be encoded are placed at positions $n-2$, $n-1$, and $n$, followed by the rest of the qubits initialized to the state $|0\rangle$. Next, the logical operators are encoded according the Algorithm \ref{alg:algo-1}. Thereafter, the stabilizers corresponding to the rows of standard form of the parity check matrix $H_{s}$ are applied according the Algorithm \ref{alg:algo-1}. The encoder circuit thus designed is shown in Fig. \ref{fig:eight-qubit-encoder}.

\begin{figure}
\begin{centering}
\includegraphics[scale=0.35]{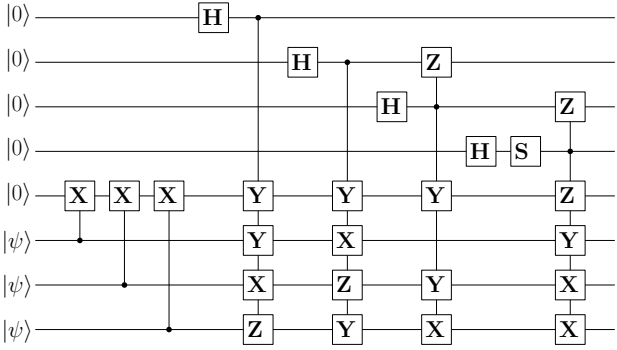}
\par\end{centering}
\caption{Modified encoder for the eight-qubit code after removing redundant controlled-$Z$ gates. \label{fig:eight-qubit-encoder-modified}}
\end{figure}

From Fig. \ref{fig:eight-qubit-encoder}, we notice that there are three controlled-$Z$ gates which act on $|0\rangle$ qubits. Since $Z$ acts trivially on $|0\rangle$, those $Z$ gates can be ignored, resulting in the modified encoding circuit as shown in Fig. \ref{fig:eight-qubit-encoder-modified}.

\subsection{Syndrome measurement circuit and error corrector}

The syndromes are unique as shown in Table \ref{tab:tab-syn-eight-qubit}.
Each qubit in the eight-qubit code can be affected by three kind of
errors, namely $X$, $Y$, and $Z$ errors. So, there are $24$ different
types of single qubit errors possible, each of which gives a different
syndrome as shown in Table \ref{tab:tab-syn-eight-qubit}. Each bit in the 5-bit syndrome represents whether the corresponding stabilizer commutes with the error. If it commutes, the bit is 0, else it is 1. It should also be observed that each syndrome
is unique as shown in Table \ref{tab:tab-syn-eight-qubit}.

\begin{table}
\caption{Syndrome table for the eight-qubit code.\label{tab:tab-syn-eight-qubit}}

\centering{}
\global\long\def\arraystretch{1.2}%
\begin{tabular}{|p{0.1cm}p{0.1cm}p{0.1cm} p{0.1cm}p{0.1cm}p{0.1cm}p{0.1cm} p{0.1cm}|p{0.25cm}|p{0.25cm}|p{0.25cm}|p{0.25cm}|p{0.25cm}|p{0.8cm}|}
\hline 
 &  &  &  &  &  &  &  & $M_{1}$ & $M_{2}$ & $M_{3}$ & $M_{4}$ & $M_{5}$ & Decimal value\tabularnewline
\hline 
\hline 
$X$ & $I$ & $I$ & $I$ & $I$ & $I$ & $I$ & $I$ & $0$ & $0$ & $0$ & $0$ & $1$ & $1$\tabularnewline
\hline 
$Z$ & $I$ & $I$ & $I$ & $I$ & $I$ & $I$ & $I$ & $1$ & $0$ & $0$ & $0$ & $0$ & $16$\tabularnewline
\hline 
$Y$ & $I$ & $I$ & $I$ & $I$ & $I$ & $I$ & $I$ & $1$ & $0$ & $0$ & $0$ & $1$ & $17$\tabularnewline
\hline 
$I$ & $X$ & $I$ & $I$ & $I$ & $I$ & $I$ & $I$ & $1$ & $0$ & $1$ & $0$ & $1$ & $21$\tabularnewline
\hline 
$I$ & $Z$ & $I$ & $I$ & $I$ & $I$ & $I$ & $I$ & $0$ & $1$ & $0$ & $0$ & $0$ & $8$\tabularnewline
\hline 
$I$ & $Y$ & $I$ & $I$ & $I$ & $I$ & $I$ & $I$ & $1$ & $1$ & $1$ & $0$ & $1$ & $29$\tabularnewline
\hline 
$I$ & $I$ & $X$ & $I$ & $I$ & $I$ & $I$ & $I$ & $0$ & $1$ & $0$ & $1$ & $1$ & $11$\tabularnewline
\hline 
$I$ & $I$ & $Z$ & $I$ & $I$ & $I$ & $I$ & $I$ & $0$ & $0$ & $1$ & $0$ & $0$ & $4$\tabularnewline
\hline 
$I$ & $I$ & $Y$ & $I$ & $I$ & $I$ & $I$ & $I$ & $0$ & $1$ & $1$ & $1$ & $1$ & $15$\tabularnewline
\hline 
$I$ & $I$ & $I$ & $X$ & $I$ & $I$ & $I$ & $I$ & $0$ & $0$ & $1$ & $1$ & $1$ & $7$\tabularnewline
\hline 
$I$ & $I$ & $I$ & $Z$ & $I$ & $I$ & $I$ & $I$ & $0$ & $0$ & $0$ & $1$ & $0$ & $2$\tabularnewline
\hline 
$I$ & $I$ & $I$ & $Y$ & $I$ & $I$ & $I$ & $I$ & $0$ & $0$ & $1$ & 0 & $1$ & $5$\tabularnewline
\hline 
$I$ & $I$ & $I$ & $I$ & $X$ & $I$ & $I$ & $I$ & $1$ & $1$ & $1$ & $1$ & $1$ & $31$\tabularnewline
\hline 
$I$ & $I$ & $I$ & $I$ & $Z$ & $I$ & $I$ & $I$ & $1$ & $1$ & $1$ & $0$ & $0$ & $28$\tabularnewline
\hline 
$I$ & $I$ & $I$ & $I$ & $Y$ & $I$ & $I$ & $I$ & $0$ & $0$ & $0$ & $1$ & $1$ & $3$\tabularnewline
\hline 
$I$ & $I$ & $I$ & $I$ & $I$ & $X$ & $I$ & $I$ & $1$ & $0$ & $0$ & $1$ & $1$ & $19$\tabularnewline
\hline 
$I$ & $I$ & $I$ & $I$ & $I$ & $Z$ & $I$ & $I$ & $1$ & $1$ & $0$ & $1$ & $0$ & $26$\tabularnewline
\hline 
$I$ & $I$ & $I$ & $I$ & $I$ & $Y$ & $I$ & $I$ & $0$ & $1$ & $0$ & $0$ & $1$ & $9$\tabularnewline
\hline 
$I$ & $I$ & $I$ & $I$ & $I$ & $I$ & $X$ & $I$ & $0$ & $1$ & $1$ & $0$ & $1$ & $13$\tabularnewline
\hline 
$I$ & $I$ & $I$ & $I$ & $I$ & $I$ & $Z$ & $I$ & $1$ & $0$ & $1$ & $1$ & $0$ & $22$\tabularnewline
\hline 
$I$ & $I$ & $I$ & $I$ & $I$ & $I$ & $Y$ & $I$ & $1$ & $1$ & $0$ & $1$ & $1$ & $27$\tabularnewline
\hline 
$I$ & $I$ & $I$ & $I$ & $I$ & $I$ & $I$ & $X$ & $1$ & $1$ & $0$ & $0$ & $1$ & $25$\tabularnewline
\hline 
$I$ & $I$ & $I$ & $I$ & $I$ & $I$ & $I$ & $Z$ & $0$ & $1$ & $1$ & $1$ & $0$ & $14$\tabularnewline
\hline 
$I$ & $I$ & $I$ & $I$ & $I$ & $I$ & $I$ & $Y$ & $1$ & $0$ & $1$ & $1$ & $1$ & $23$\tabularnewline
\hline 
$I$ & $I$ & $I$ & $I$ & $I$ & $I$ & $I$ & $I$ & $0$ & $0$ & $0$ & $0$ & $0$ & $0$\tabularnewline
\hline 
\end{tabular}
\end{table}

\begin{figure}
\begin{centering}
\includegraphics[scale=0.33]{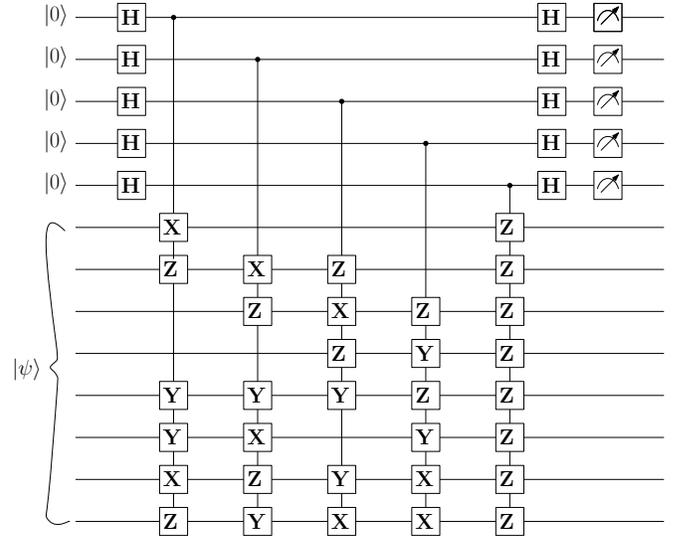}
\par\end{centering}
\caption{Syndrome measurement circuit for the eight-qubit code. \label{fig:syn-measure-1}}
\end{figure}

The syndrome measurement circuit is shown in Fig. \ref{fig:syn-measure-1}.
Five ancilla qubits are used to measure each of the six stabilizers.
Measurement of the ancilla qubits gives the syndrome. Depending on
the syndrome, appropriate error correction can be performed by using
suitable $X$, $Z$, or $Y$ gate on the appropriate qubit. A syndrome
measurement of $00000$ implies that no error has occurred. It should
also be noted that any $5$ bit syndrome other than the syndromes
mentioned in Table \ref{tab:tab-syn-eight-qubit} signifies more than
a single qubit error which cannot be corrected.

The eight-qubit code encodes 3 logical qubits to 8 physical qubits and thus has a rate of $3/8$. To have a higher coding rate, one way is to paste a five-qubit code and an eight-qubit code to get a 13-qubit code. This code encodes 7 logical qubits to 13 physical qubits, and thus has a code rate of $7/13$. It protects a single qubit from bit flip and phase flip errors. Such a code was discussed in \cite{gottesman-thesis}. We will discuss the construction, encoding and decoding circuits in the next section.

\section{Optimization of the eight-qubit encoder circuit}

The encoding circuit designed in Fig. \ref{fig:eight-qubit-encoder-modified} uses 20 2-qubit gates and 5 single qubit gates. It requires 3 different types of 2-qubit gates, the CX, CY, and CZ gates. However, for most practical purposes, only a single type of 2-qubit gate may be available. Our goal is to optimize the circuit, such that it uses CNOT gates and $H$ gates only. Since multiple qubit operations are a source of noise and decoherence, we also intend to minimize the number of CNOT gates. Our approach to optimize the circuit consists of two steps. First, we use equivalence rules related to quantum gates for conversion between gates, and for moving the gates around in the circuit. Most of these rules are applied by visual inspection of the circuit. We next use matrix equivalence to optimize a set of CNOT gates acting sequentially on a number of qubits. These steps are described in detail in the next sub-sections.

\subsection{Optimization using equivalence rules}

\begin{figure*} 
\begin{centering}
\includegraphics[scale=0.34]{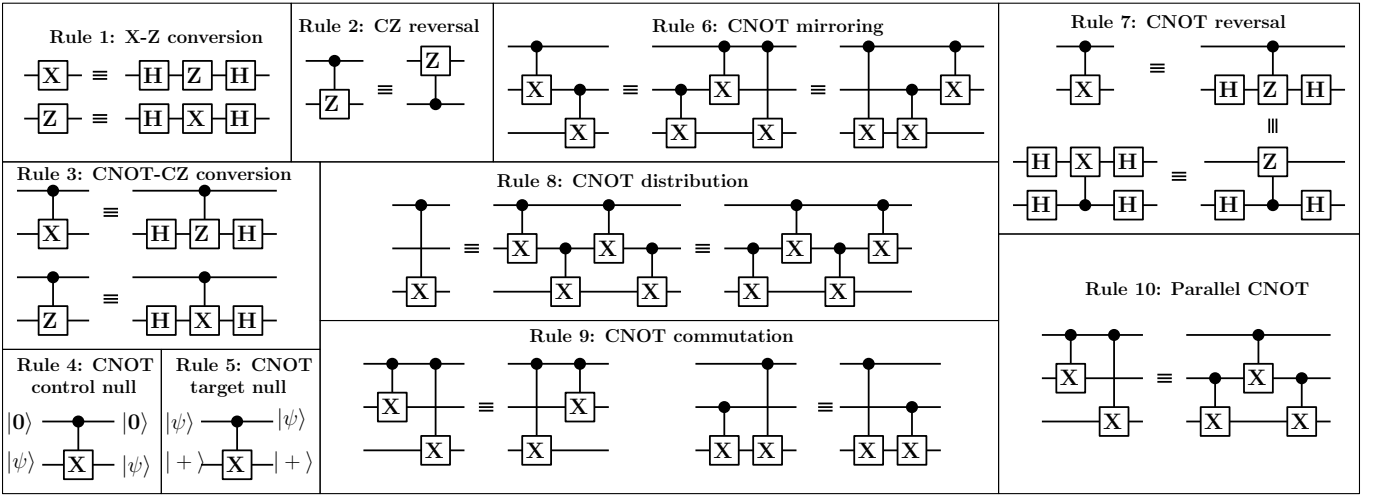}
\par\end{centering}
\centering{}\caption{Equivalence rules related to quantum circuits (based on \cite{garcia}).}
\label{fig:eq-rules}
\end{figure*}

Various equivalence rules related to quantum circuits have been discussed in \cite{garcia}. These rules are illustrated in Fig. \ref{fig:eq-rules}. Rules 1-3 relate to conversions between $X$ and $Z$ gates (or CNOT and CZ gates for 2-qubit gates). These rules also help a circuit designer to switch control and target. Rules 4-5 are useful for circuits having multiple ancilla qubits at state $|0\rangle$, or state $|+\rangle$ in the Hadamard basis. Rules 6-10 are useful for manipulation of quantum circuits, enabling gates to move past each other.

\begin{figure}
\begin{centering}
\includegraphics[scale=0.29]{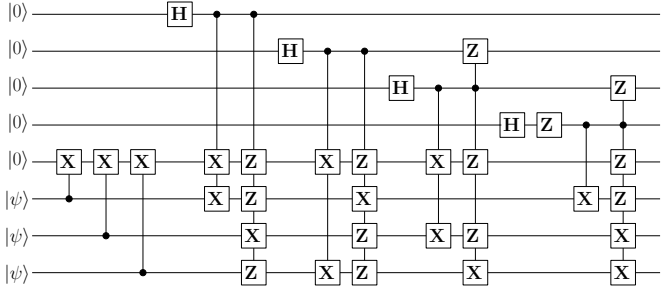}
\par\end{centering}
\caption{Encoder for the eight-qubit code using controlled-XZ gates instead of controlled-Y gates \label{fig:eight-qubit-encoder-modified-xz}}
\end{figure}

We now illustrate how the above rules can be used to find an equivalent circuit for the 8-qubit encoder shown in Fig. \ref{fig:eight-qubit-encoder-modified}. First, we represent the encoder circuit using controlled-XZ gates instead of controlled-Y gates as shown in Fig. \ref{fig:eight-qubit-encoder-modified-xz}. 

\begin{figure}
\begin{centering}
\includegraphics[scale=0.25]{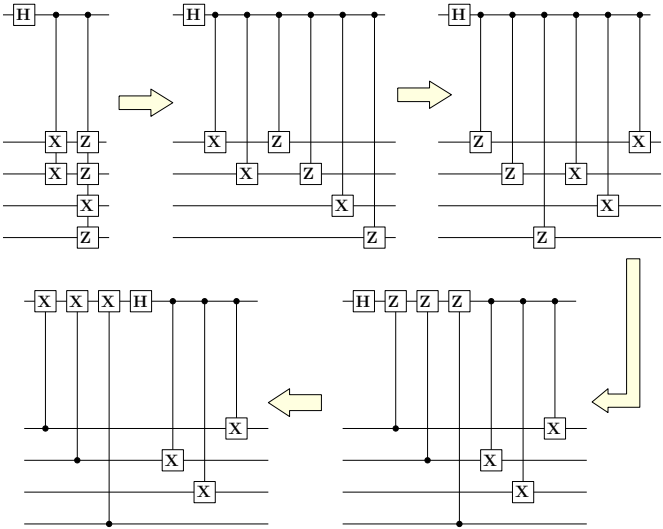}
\par\end{centering}
\caption{Conversion of CZ gates to CX gates for the first stabilizer, keeping the number of $H$ gates constant. \label{fig:eight-qubit-encoder-opt-step-1}}
\end{figure}

\begin{figure*}
\begin{centering}
\includegraphics[scale=0.32]{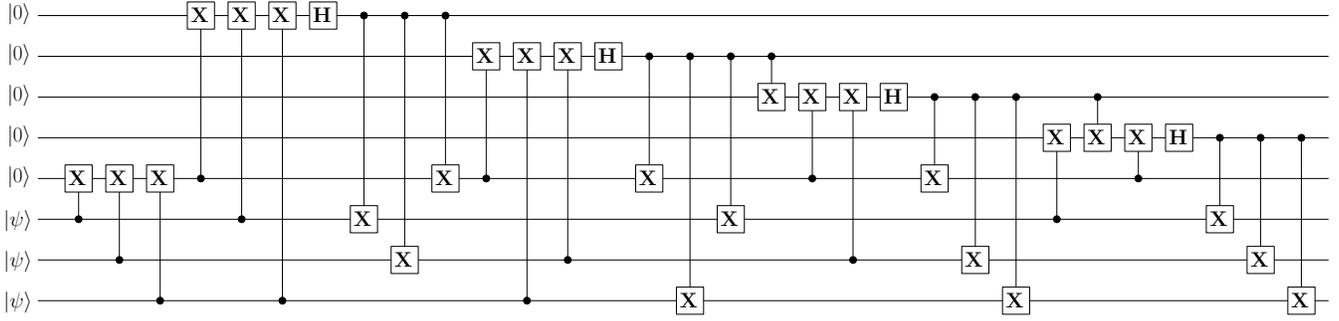}
\par\end{centering}
\caption{Encoder for the eight-qubit code after converting CZ to CNOT gates. \label{fig:eight-qubit-encoder-opt-step-2}}
\end{figure*}

Since most practical quantum computing systems use one type of 2-qubit gates, we will convert all controlled-Z gates to controlled-X gates using Rule 3 in Fig. \ref{fig:eq-rules}. However, this step will result in the increase of $H$ gates significantly. Thus, we propose a systematic way where each stabilizer is considered separately in a sequential fashion to achieve this goal. The steps for the first stabilizer is illustrated in Fig. \ref{fig:eight-qubit-encoder-opt-step-1}. First, the CZ gates are brought to the left, keeping in mind that it can commute with CX gates only in pairs to preserve the eigen value of +1. Next, control and target of the CZ gates are reversed using Rule 2. Subsequently, the CZ gates are converted to CX gates using Rule 3. Similar steps can be applied to all the stabilizers to obtain the circuit shown in Fig. \ref{fig:eight-qubit-encoder-opt-step-2}.

\begin{figure*}
\begin{centering}
\includegraphics[scale=0.35]{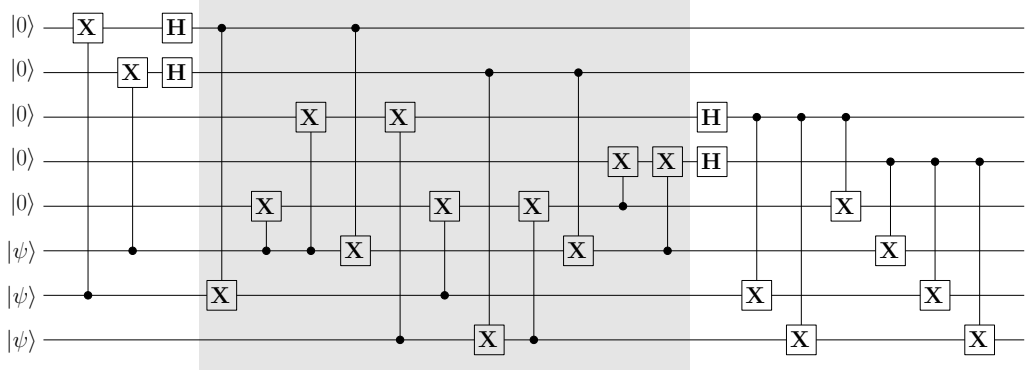}
\par\end{centering}
\caption{Encoder for the eight-qubit code after optimization using equivalence rules in Fig. \ref{fig:eq-rules}. \label{fig:eight-qubit-encoder-opt-step-3}}
\end{figure*}

Next, we use Rules 6-10 to move around the CNOT gates along with $H$ gates with the goal of reducing the number of overall CNOT gates. The encoder circuit has four ancilla qubits set to $|0\rangle$. Noting that according to Rule 6, a CNOT with control set to $|0\rangle$ is nullified, we move some CNOT gates in such a way that their controls are brought to the $|0\rangle$ ancillas, thereby annihilating those. Using the above steps, we could optimize the circuit to 19 CNOT and 4 Hadamard gates as shown in Fig. \ref{fig:eight-qubit-encoder-opt-step-3}.

\subsection{Optimization using group theoretic matrix equivalence}

After we have optimized the circuit using the equivalence rules, it may still be possible to optimize the circuit further. However, it may be difficult to do so using visual inspection. To accomplish this goal, we will use some approaches described in \cite{bataille2022quantum}. We show an example of matrix equivalence for quantum circuits containing CNOT gates as described next.

Consider an $n$-qubit quantum circuit consisting of only CNOT gates. Let the initial state of the circuit be represented by an $n\times n$ identity matrix. A CNOT gate acting between qubits $k_{1}$ and $k_{2}$ with control on $k_{1}$ and target on $k_{2}$ can be represented by the row transformation $R_{k_{2}}\rightarrow R_{k_{1}}+R_{k_{2}}$. Thus, the entire circuit can be represented by a series of elementary row transformations. Two $n$-qubit quantum circuits are equivalent if they result in the same matrix after their respective elementary
row transformations. For example, let us consider Rule 8 (CNOT distribution) in Fig. \ref{fig:eq-rules}.

The leftmost circuit consists of a single elementary row transformation $R_{3}\rightarrow R_{1}+R_{3}$, which results in

\[
\left[\begin{array}{ccc}
1 & 0 & 0\\
0 & 1 & 0\\
1 & 0 & 1
\end{array}\right]
\]

The middle circuit consists of four row transformations $R_{2}\rightarrow R_{1}+R_{2}$,
$R_{3}\rightarrow R_{2}+R_{3}$, $R_{2}\rightarrow R_{1}+R_{2}$,
and $R_{3}\rightarrow R_{2}+R_{3}$. This results in 
\[
\left[\begin{array}{ccc}
1 & 0 & 0\\
0 & 1 & 0\\
1 & 0 & 1
\end{array}\right]
\]

The rightmost circuit consists of four row transformations $R_{3}\rightarrow R_{2}+R_{3}$,
$R_{2}\rightarrow R_{1}+R_{2}$, $R_{3}\rightarrow R_{2}+R_{3}$,
and $R_{2}\rightarrow R_{1}+R_{2}$. This results in 
\[
\left[\begin{array}{ccc}
1 & 0 & 0\\
0 & 1 & 0\\
1 & 0 & 1
\end{array}\right]
\]

We observe that all the three circuits result in the same final matrix, and are thus equivalent. The authors in  \cite{bataille2022quantum} note that for a $n$-qubit quantum circuit consisting of only CNOT gates, the input-output transformation of states can be represented by a series of elementary row transformations. Taking the final matrix after the transformations, and applying a Gaussian elimination method to that leads to a set of steps resulting in the identity matrix. These sets of steps applied in reverse to the identity matrix leads to an equivalent circuit for the initial circuit. However, this circuit may not be optimum. We will try to apply this procedure to the shaded region in Fig. \ref{fig:eight-qubit-encoder-opt-step-3}, which consists of $11$ CNOT gates. We start with a $8\times 8$ identity matrix $I$, and applying the $11$ sequential row transformations as follows:

\begin{align} \label{eq:orig-trf}
R_{7}&\rightarrow R_{1}+R_{7}\nonumber\\
R_{5}&\rightarrow R_{5}+R_{6}\nonumber\\
R_{3}&\rightarrow R_{3}+R_{6}\nonumber\\
R_{6}&\rightarrow R_{6}+R_{1}\nonumber\\
R_{3}&\rightarrow R_{3}+R_{8}\nonumber\\
R_{5}&\rightarrow R_{5}+R_{7}\nonumber\\
R_{8}&\rightarrow R_{8}+R_{2}\nonumber\\
R_{5}&\rightarrow R_{5}+R_{8}\nonumber\\
R_{6}&\rightarrow R_{6}+R_{2}\nonumber\\ 
R_{4}&\rightarrow R_{4}+R_{5}\nonumber\\
R_{4}&\rightarrow R_{4}+R_{6}
\end{align}

At the end of the row transformations, we obtain the matrix

\begin{equation}
T=
\left[\begin{array}{cccccccc}
1 & 0 & 0 & 0 & 0 & 0 & 0 & 0\\
0 & 1 & 0 & 0 & 0 & 0 & 0 & 0\\
0 & 0 & 1 & 0 & 0 & 1 & 0 & 1\\
0 & 0 & 0 & 1 & 1 & 0 & 1 & 1\\
1 & 1 & 0 & 0 & 1 & 1 & 1 & 1\\
1 & 1 & 0 & 0 & 0 & 1 & 0 & 0\\
1 & 0 & 0 & 0 & 0 & 0 & 1 & 0\\
0 & 1 & 0 & 0 & 0 & 0 & 0 & 1
\end{array}\right]
\end{equation}

Gaussian elimination on $T$ requires $14$ steps (equivalent to $14$ CNOT gates), which is less optimized than the original circuit. However, applying the following steps sequentially to $T$ gives the $8\times 8$ identity matrix.

\begin{align}\label{eq:opt-trf}
R_{5}&\rightarrow R_{5}+R_{6}\nonumber\\
R_{4}&\rightarrow R_{4}+R_{5}\nonumber\\
R_{7}&\rightarrow R_{7}+R_{1}\nonumber\\
R_{8}&\rightarrow R_{8}+R_{2}\nonumber\\
R_{6}&\rightarrow R_{6}+R_{1}\nonumber\\
R_{6}&\rightarrow R_{6}+R_{2}\nonumber\\
R_{3}&\rightarrow R_{3}+R_{6}\nonumber\\
R_{3}&\rightarrow R_{3}+R_{8}\nonumber\\
R_{5}&\rightarrow R_{5}+R_{7}\nonumber\\
R_{5}&\rightarrow R_{5}+R_{8}
\end{align}

\begin{figure*}
\begin{centering}
\includegraphics[scale=0.35]{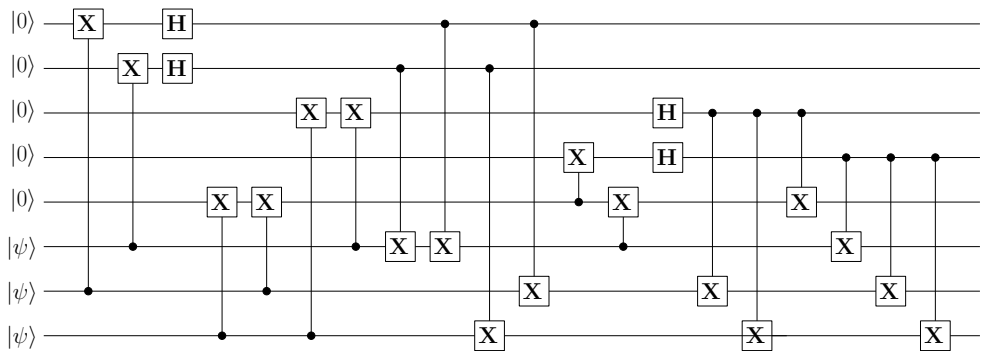}
\par\end{centering}
\caption{Final optimized encoding circuit for the eight-qubit code after optimizing shaded region in Fig. \ref{fig:eight-qubit-encoder-opt-step-3} using matrix equivalence. \label{fig:eight-qubit-encoder-opt-step-4}}
\end{figure*}

Thus, the operations in equation (\ref{eq:opt-trf}) applied in reverse is equivalent to the operations in equation (\ref{eq:orig-trf}). This is more optimized since it has $10$ transformations, and thus corresponds to $10$ CNOT gates. The resulting encoding circuit 
consisting of $18$ CNOT and $4$ $H$ gates is shown in Fig. \ref{fig:eight-qubit-encoder-opt-step-4}.

\section{Optimized encoder circuit design for Steane code and a 13-qubit code}

Steane code \cite{steane1996multiple} is a CSS code which uses the Hamming
$[7,4,3]$ code and the dual of the Hamming code, i.e., the $[7,3,4]$
code to correct bit flip and phase flip errors respectively. The $[7,4,3]$
Hamming code contains its dual, and thus can be used in the CSS framework
to obtain a quantum ECC. One logical qubit is encoded into seven physical
qubits, thus enabling the Steane code to detect and correct a single
qubit error. In stabilizer framework, the Steane code is represented
by six generators as shown below:
\begin{center}
\begin{tabular}{c|ccccccc}
$M_{1}$  & $X$  & $X$  & $X$  & $X$  & $I$  & $I$  & $I$\tabularnewline
$M_{2}$  & $X$  & $X$  & $I$  & $I$  & $X$  & $X$  & $I$\tabularnewline
$M_{3}$  & $X$  & $I$  & $X$  & $I$  & $X$  & $I$  & $X$\tabularnewline
$M_{4}$  & $Z$  & $Z$  & $Z$  & $Z$  & $I$  & $I$  & $I$\tabularnewline
$M_{5}$  & $Z$  & $Z$  & $I$  & $I$  & $Z$  & $Z$  & $I$\tabularnewline
$M_{6}$  & $Z$  & $I$  & $Z$  & $I$  & $Z$  & $I$  & $Z$\tabularnewline
\end{tabular}. 
\par\end{center}

\begin{figure*}
\begin{centering}
\includegraphics[scale=0.35]{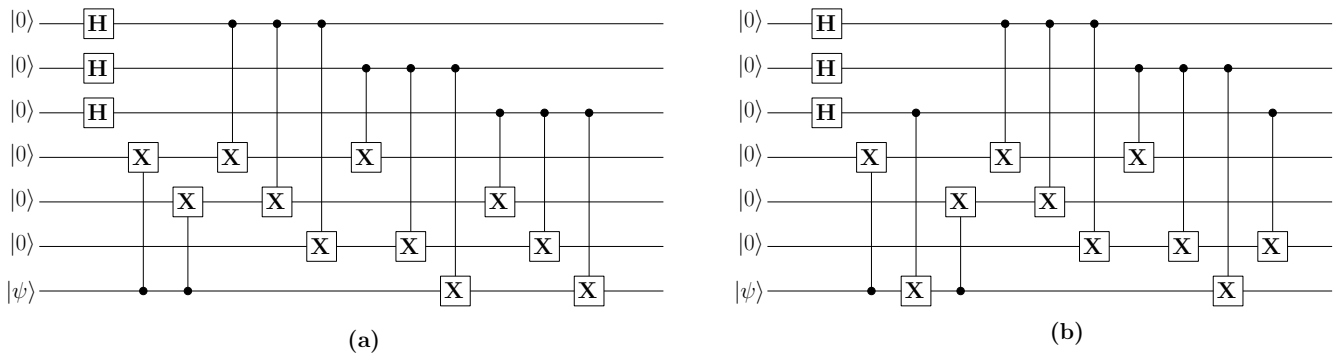}
\par\end{centering}
\caption{(a) Encoder for Steane code using Algorithm \ref{alg:algo-1}. (b) Optimized encoder for the Steane code using equivalence rules. \label{fig:steane-encoder-opt}}
\end{figure*}

We designed a Steane code encoder using Algorithm \ref{alg:algo-1} as shown in Fig. \ref{fig:steane-encoder-opt} (a). Using the equivalence rules the total number of CNOT gates was reduced by 1 as shown in Fig. \ref{fig:steane-encoder-opt} (b).

New stabilizer codes can be generated by merging available stabilizer codes. Such a code can be created by pasting a five-qubit code [[5,1,3]] and an eight-qubit code [[8,3,3]] has been discussed in \cite{gottesman-thesis}. The new 13-qubit code [[13,7,3]] has a rate of $7/13$. It can be represented by the following stabilizers:

\begin{center}
\scalebox{0.85}{
\begin{tabular}{c|ccccccccccccc}
$M_{1}$ & $X$ & $X$ & $X$ & $X$ & $X$ & $X$ & $X$ & $X$ & $I$ & $I$ & $I$ & $I$ & $I$\tabularnewline
$M_{2}$ & $Z$ & $Z$ & $Z$ & $Z$ & $Z$ & $Z$ & $Z$ & $Z$ & $I$ & $I$ & $I$ & $I$ & $I$\tabularnewline
$M_{3}$ & $I$ & $I$ & $I$ & $I$ & $I$ & $I$ & $I$ & $I$ & $X$ & $Z$ & $Z$ & $X$ & $I$\tabularnewline
$M_{4}$ & $I$ & $X$ & $I$ & $X$ & $Y$ & $Z$ & $Y$ & $Z$ & $I$ & $X$ & $Z$ & $Z$ & $X$\tabularnewline
$M_{5}$ & $I$ & $X$ & $Z$ & $Y$ & $I$ & $X$ & $Z$ & $Y$ & $X$ & $I$ & $X$ & $Z$ & $Z$\tabularnewline
$M_{6}$ & $I$ & $Y$ & $X$ & $Z$ & $X$ & $Z$ & $I$ & $Y$ & $Z$ & $X$ & $I$ & $X$ & $Z$\tabularnewline
\end{tabular}. }
\par\end{center}

\begin{figure}
\begin{centering}
\includegraphics[scale=0.25]{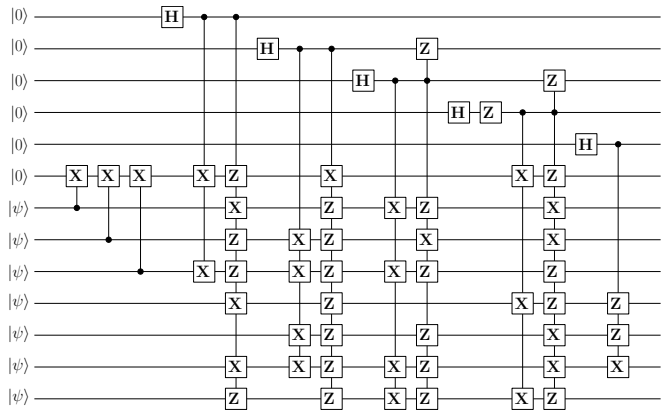}
\par\end{centering}
\caption{13-qubit encoder designed using Algorithm \ref{alg:algo-1}. CNOT and CZ gates were used as 2-qubit gates. \label{fig:13-qubit-encoder-xz}}
\end{figure}

Using Algorithm \ref{alg:algo-1}, the encoder circuit for the 13-qubit code was designed using CNOT and CZ gates as shown in Fig. \ref{fig:13-qubit-encoder-xz}. It uses 50 2-qubit gates and 5 single qubit gates.

\begin{figure*}
\begin{centering}
\includegraphics[scale=0.22]{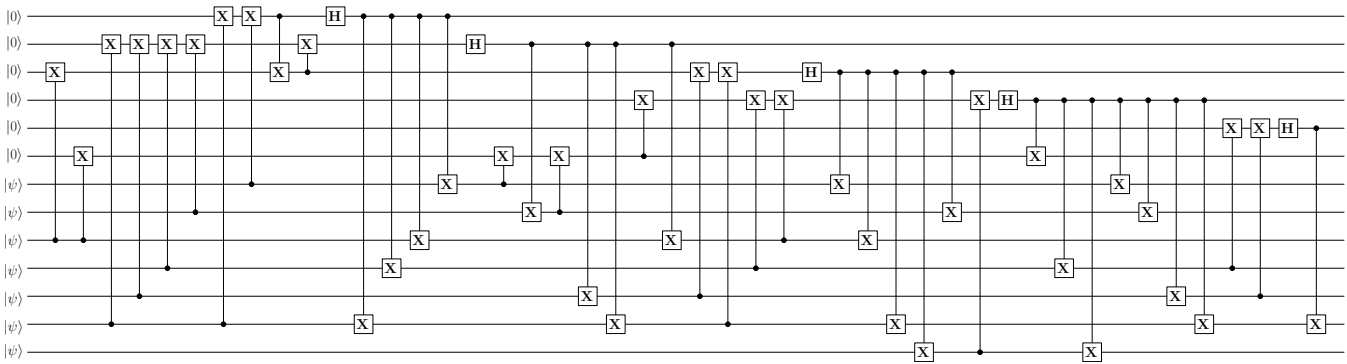}
\par\end{centering}
\caption{Optimized 13-qubit encoder using 41 CNOT gates and 5 $H$ gates. \label{fig:13-qubit-encoder-opt}}
\end{figure*}

We also optimized the encoding circuit as described in Section IV to reduce the number of gates to 41 CNOT gates and 5 $H$ gates. The resulting circuit is presented in Fig. \ref{fig:13-qubit-encoder-opt}.

\section{Results}

The combined encoder and decoder circuits for the 8-qubit code, Steane code, and the 13-qubit code were simulated using IBM Qiskit. Circuits designed from Algorithm \ref{alg:algo-1} as well as the optmized circuits derived from those were simulated using IBM Qiskit. Errors were introduced at different positions to test for correctability. Syndromes were found to match with Table \ref{tab:tab-syn-eight-qubit} for the 8-qubit code. The syndrome table for the Steane code and the 13-qubit code has not been included in this paper for lack of space. However, those tables can be derived using the same procedure as the 8-qubit code. The syndromes were verified to match with the respective tables.

Another important parameter to measure the efficiency of the quantum
circuits is the number of single and multiple qubit gates used by
the quantum circuits. We list the number of gates used in the quantum circuits presented in this paper in
Table \ref{tab:Resource-utilization}. In the second column, we observe that the eight-qubit encoder designed using Algorithm \ref{alg:algo-1} requires 8 CNOT gates, 7 CY gates, and 5 CZ gates, thus using a total of 20 2-qubit gates. In practical applications, one may need to use a combination of CZ and CNOT gates to obtain CY gates. The third column shows the number of gates used in a circuit using CZ and CNOT gates. We observe that it used a total of 27 2-qubit gates (15 CNOT and 12 CZ). For practical considerations, only one type of 2-qubit gate may be available. To achieve this, the circuit was further optimized using the equivalence rules to reduce the number of gates to $18$ CNOT gates and 4 $H$ gates, as reported in column 6 of Table \ref{tab:Resource-utilization}. The reduction in 2-qubit gates from $27$ gates to $18$ gates is quite significant\footnote{It should be noted that CY gates are usually designed as a combination of CNOT and CZ gates. Thus, a single CY gate is  equivalent to two 2-qubit gates.}. Columns 5 and 6 show the number of gates used in Steane code \cite{steane1996multiple} encoder and its optimized version. The number of CNOT gates is reduced from 11 to 10. We also report the number of gates used in the 13-qubit encoder and its optimized version. The proposed approach achieves a 20 \% reduction in the number of 2-qubit gates, from 26 CNOT and 24 CZ gates to 41 CNOT gates.

In \cite{dong2013efficient}, an encoding rule was formulated by observing all the $2^k$ different codewords for the eight-qubit code. For quantum codes where $k$ is significantly larger, the method in \cite{dong2013efficient} would be very complex because one has to figure out an encoding rule by observing $2^k$ codewords. Additionally, the method in \cite{dong2013efficient} requires more than twice the number of gates than the eight-qubit code encoder circuit proposed in this paper. Furthermore, the circuit in \cite{dong2013efficient} uses an array of different types of multi-qubit gates which may not be desirable for practical considerations. Compared to \cite{dong2013efficient} which requires 14 single qubit gates, 33 2-qubit gates, and 6 3-qubit gates, the proposed eight-qubit encoder requires only 18 CNOT gates and 4 $H$ gates. The syndrome measurement circuit in this paper requires slightly less number of gates than \cite{dong2013efficient}.

\begin{table*}

\caption{Resource utilization summary for the various designed quantum circuits
in terms of number of gates used. \label{tab:Resource-utilization}}

\begin{centering}
\global\long\def\arraystretch{1.5}%
\begin{tabular}{|>{\centering}p{5cm}|>{\centering}p{1cm}|>{\centering}p{1cm}|>{\centering}p{1cm}|>{\centering}p{1cm}|>{\centering}p{1cm}|>{\centering}p{1cm}|>{\centering}p{1cm}|>{\centering}p{1cm}|}
\hline 
\rowcolor{Gray}
Parameters & $H$ gates & $X$ gates & $S$ gates & $Z$ gates & CNOT gates & CY gates & CZ gates & CCNOT gates\tabularnewline
\hline 
\rowcolor{LightCyan}
Proposed eight qubit encoder (using CNOT, CZ, and CY gates)  & 4 & 0 & 1 & 0 & 8 & 7 & 5 & 0\tabularnewline
\hline 
\rowcolor{LightCyan}
Proposed eight qubit encoder (using CNOT and CZ gates) & 4 & 0 & 0 & 1 & 15 & 0 & 12 & 0\tabularnewline
\hline 
\rowcolor{LightCyan}
\textbf{Proposed eight qubit encoder (optimized using equivalence rules and
matrix equivalence)} & \textbf{4} & \textbf{0} & \textbf{0} & \textbf{0} & \textbf{18} & \textbf{0} & \textbf{0} & \textbf{0}\tabularnewline
\hline 
\rowcolor{LightCyan}
Eight qubit encoder \cite{dong2013efficient}  & 4 & 10 & 0 & 0 & 24 & 0 & 9 & 6\tabularnewline
\hline 
\rowcolor{LightCyan}
\textbf{Proposed eight qubit syndrome measurement} & \textbf{10} & \textbf{0} & \textbf{0} & \textbf{0} & \textbf{8} & \textbf{8} & \textbf{16} & \textbf{0}\tabularnewline
\hline 
\rowcolor{LightCyan}
Eight qubit syndrome measurement \cite{dong2013efficient}  & 10 & 0 & 0 & 0 & 14 & 6 & 14 & 0\tabularnewline
\hline
\rowcolor{LightGreen}
Steane code encoder & 3 & 0 & 0 & 0 & 11 & 0 & 0 & 0\tabularnewline
\hline 
\rowcolor{LightGreen}
\textbf{Optimized Steane code encoder} & \textbf{3} & \textbf{0} & \textbf{0} & \textbf{0} & \textbf{10} & \textbf{0} & \textbf{0} & \textbf{0}\tabularnewline
\hline 
\rowcolor{LightYellow}
13-qubit code encoder & 5 & 0 & 0 & 1 & 26 & 0 & 24 & 0\tabularnewline
\hline 
\rowcolor{LightYellow}
\textbf{Optimized 13-qubit code encoder} & \textbf{5} & \textbf{0} & \textbf{0} & \textbf{0} & \textbf{41} & \textbf{0} & \textbf{0} & \textbf{0}\tabularnewline
\hline

\end{tabular}
\par\end{centering}
\end{table*}

\section{Conclusions}

This paper has presented a detailed procedure for construction of encoder and decoder
circuits for stabilizer codes. Further, we optimized of the circuits in terms of the number of gates used. The procedure was used to design optimized encoder and decoder circuits for an eight-qubit code. The circuits were verified using IBM Qiskit. The encoding circuits designed using the proposed procedure result in a reduction in the number of gates by a factor of 2 for the eight-qubit code, compared to prior designs. We also designed optimized encoder circuits for Steane code and a 13-qubit code. For larger code lengths involving significantly higher number of qubits, it is difficult to optimize the circuits manually. Therefore, future efforts will be directed towards developing an automated tool to optimize the quantum circuits. Designing encoder-decoder circuits for EA stabilizer codes is also a topic of further research. Additionally, we plan to use the methods in this paper towards design of quantum circuits for more complex quantum ECCs such as BCH, LDPC, and polar codes.

\begin{appendices}

\section{Mathematical verification of Algorithm \ref{alg:algo-1} through eight-qubit code}

In this Appendix, we verify that Algorithm \ref{alg:algo-1} results in the same state as given by equation (\ref{eq:encoded-states}) for the eight-qubit code. Out of the eight possible states, we only verify the application of the algorithm for the state $\overline{000}$. The algorithm can be verified for other states using suitable $\overline{X}$ operators.
operators

From equation (\ref{eq:encoded-states}) we have,

\begin{align}
|\overline{000}\rangle= & (I+M_{1})(I+M_{2})(I+M_{3})(I+M_{4})|00000000\rangle
\end{align}

Expanding, we get

\begin{align}
|\overline{000}\rangle= & |00000000\rangle+M_{1}|00000000\rangle+M_{2}|00000000\rangle\nonumber \\
 & +M_{3}|00000000\rangle+M_{4}|00000000\rangle+M_{1}M_{2}|00000000\rangle\nonumber \\
 & +M_{1}M_{3}|00000000\rangle+M_{1}M_{4}|00000000\rangle\nonumber \\
 & +M_{2}M_{3}|00000000\rangle+M_{2}M_{4}|00000000\rangle\nonumber \\
 & +M_{3}M_{4}|00000000\rangle+M_{1}M_{2}M_{3}|00000000\rangle\nonumber \\
 & +M_{1}M_{2}M_{4}|00000000\rangle+M_{1}M_{3}M_{4}|00000000\rangle\nonumber \\
 & +M_{2}M_{3}M_{4}|00000000\rangle+M_{1}M_{2}M_{3}M_{4}|00000000\rangle\\
 =& \frac{1}{4}(|00000000\rangle-|00010111\rangle-|00101011\rangle\nonumber \\
 & +|00111100\rangle-|01001101\rangle+|01011010\rangle\nonumber \\
 & +|01100110\rangle-|01110001\rangle-|10001110\rangle\nonumber \\
 & +|10011001\rangle+|10100101\rangle-|10110010\rangle\nonumber \\
 & +|11000011\rangle-|11010100\rangle-|11101000\rangle\nonumber \\
 & +|11111111\rangle) \label{eq:000-state}
\end{align}

We now verify that Algorithm \ref{alg:algo-1}, and the circuit in Fig. \ref{fig:eight-qubit-encoder} result in the same state as in equation (\ref{eq:000-state}).
We have the initial state $\psi=|00000000\rangle$. 

\textbf{Step A:} Applying $H$ gate on qubit $1$, we have

\begin{align}
|\psi_1\rangle= & \frac{1}{\sqrt{2}} (|00000000\rangle+|10000000\rangle)
\end{align}

\textbf{Step B:} Applying $M_1$ controlled at qubit $1$ we have,

\begin{align}
|\psi_2\rangle= & \frac{1}{\sqrt{2}} (|00000000\rangle+(i\cdot i)|10001110\rangle)
\nonumber \\
=& \frac{1}{\sqrt{2}} (|00000000\rangle -|10001110\rangle)
\end{align}

\textbf{Step C:} Applying $H$ gate on qubit $2$, we have

\begin{align}
|\psi_3\rangle= 
& \frac{1}{2} (|00000000\rangle + |01000000\rangle -|10001110\rangle
\nonumber \\
& -|11001110\rangle)
\end{align}

\textbf{Step D:} Applying $M_2$ controlled at qubit $2$ we have,

\begin{align}
|\psi_4\rangle= 
& \frac{1}{2} (|00000000\rangle + (i\cdot i)|01001101\rangle -|10001110\rangle
\nonumber \\
& - (-i\cdot -1 \cdot i)|11000011\rangle)
\nonumber \\
=& \frac{1}{2} (|00000000\rangle - |01001101\rangle -|10001110\rangle
\nonumber \\
& +|11000011\rangle)
\end{align}

\textbf{Step E:} Applying $H$ gate on qubit $3$, we have

\begin{align}
|\psi_5\rangle= 
& \frac{1}{2\sqrt{2}} (|00000000\rangle + |00100000\rangle - |01001101\rangle 
\nonumber \\
&-|01101101\rangle -|10001110\rangle - |10101110\rangle
\nonumber \\
& +|11000011\rangle +|11100011\rangle)
\end{align}

\textbf{Step F:} Applying $M_3$ controlled at qubit $3$ we have,

\begin{align}
|\psi_6\rangle= 
& \frac{1}{2\sqrt{2}} (|00000000\rangle + (i\cdot i)|00101011\rangle - |01001101\rangle 
\nonumber \\
&-(-1\cdot -i\cdot i)|01100110\rangle -|10001110\rangle 
\nonumber \\
& -(-i\cdot -i)|10100101\rangle +|11000011\rangle 
\nonumber \\
& +(-1\cdot i\cdot -i)|11101000\rangle)
\nonumber \\
=& \frac{1}{2\sqrt{2}} (|00000000\rangle -|00101011\rangle - |01001101\rangle 
\nonumber \\
&+|01100110\rangle -|10001110\rangle 
+|10100101\rangle 
\nonumber \\
&+|11000011\rangle 
-|11101000\rangle)
\end{align}

\textbf{Step G:} Applying $H$ gate followed by $S$ gate on qubit $4$, we have

\begin{align}
|\psi_7\rangle= 
& \frac{1}{4} (|00000000\rangle + i|00010000\rangle -|00101011\rangle 
\nonumber \\
&-i|00111011\rangle -|01001101\rangle 
-i|01011101\rangle 
\nonumber \\
&+|01100110\rangle +i|01110110\rangle -|10001110\rangle 
\nonumber \\
&-i|10011110\rangle +|10100101\rangle +i|10110101\rangle 
\nonumber \\
&+|11000011\rangle +i|11010011\rangle 
-|11101000\rangle
\nonumber \\
&-i|11111000\rangle)
\end{align}

\textbf{Step H:} Applying $M_4$ controlled at qubit $4$ we have,

\begin{align}
|\psi_8\rangle= 
& \frac{1}{4} (|00000000\rangle + (i\cdot i)|00010111\rangle -|00101011\rangle 
\nonumber \\
&(-i\cdot -1\cdot -1\cdot i)|00111100\rangle -|01001101\rangle 
\nonumber \\
&-(i\cdot -1\cdot -i)|01011010\rangle 
+|01100110\rangle 
\nonumber \\
&+(i\cdot -1\cdot -i)|01110001\rangle -|10001110\rangle 
\nonumber \\
&-(i\cdot -1\cdot -i)|10011001\rangle +|10100101\rangle 
\nonumber \\
&+(i\cdot -1\cdot -i)|10110010\rangle 
+|11000011\rangle 
\nonumber \\
&+(i\cdot i)|11010100\rangle 
-|11101000\rangle
\nonumber \\
&-(i\cdot -1\cdot -1\cdot i)|11111111\rangle)
\nonumber \\
=& \frac{1}{4} (|00000000\rangle -
|00010111\rangle -|00101011\rangle 
\nonumber \\
&+|00111100\rangle -|01001101\rangle 
+|01011010\rangle 
\nonumber \\
&+|01100110\rangle -|01110001\rangle -|10001110\rangle 
\nonumber \\
&+|10011001\rangle +|10100101\rangle -|10110010\rangle 
\nonumber \\
&+|11000011\rangle -|11010100\rangle 
-|11101000\rangle
\nonumber \\
&+|11111111\rangle) \label{eq:psi-8}
\end{align}

We can clearly observe that the right hand side of equation (\ref{eq:000-state}) is same as the equation (\ref{eq:psi-8}). Thus, $|\psi_8\rangle=|\overline{000}\rangle$, implying that the final state given by Algorithm \ref{alg:algo-1} or Fig. \ref{fig:eight-qubit-encoder} is $|\overline{000}\rangle$. 
The algorithm can be verified for the remaining states by applying suitable $\overline{X}$ operators to state $|\overline{000}\rangle$. 

\end{appendices}

\bibliographystyle{IEEEtran}
\bibliography{references}

\begin{IEEEbiography}[{\includegraphics[width=1in,height=1.25in,clip,keepaspectratio]{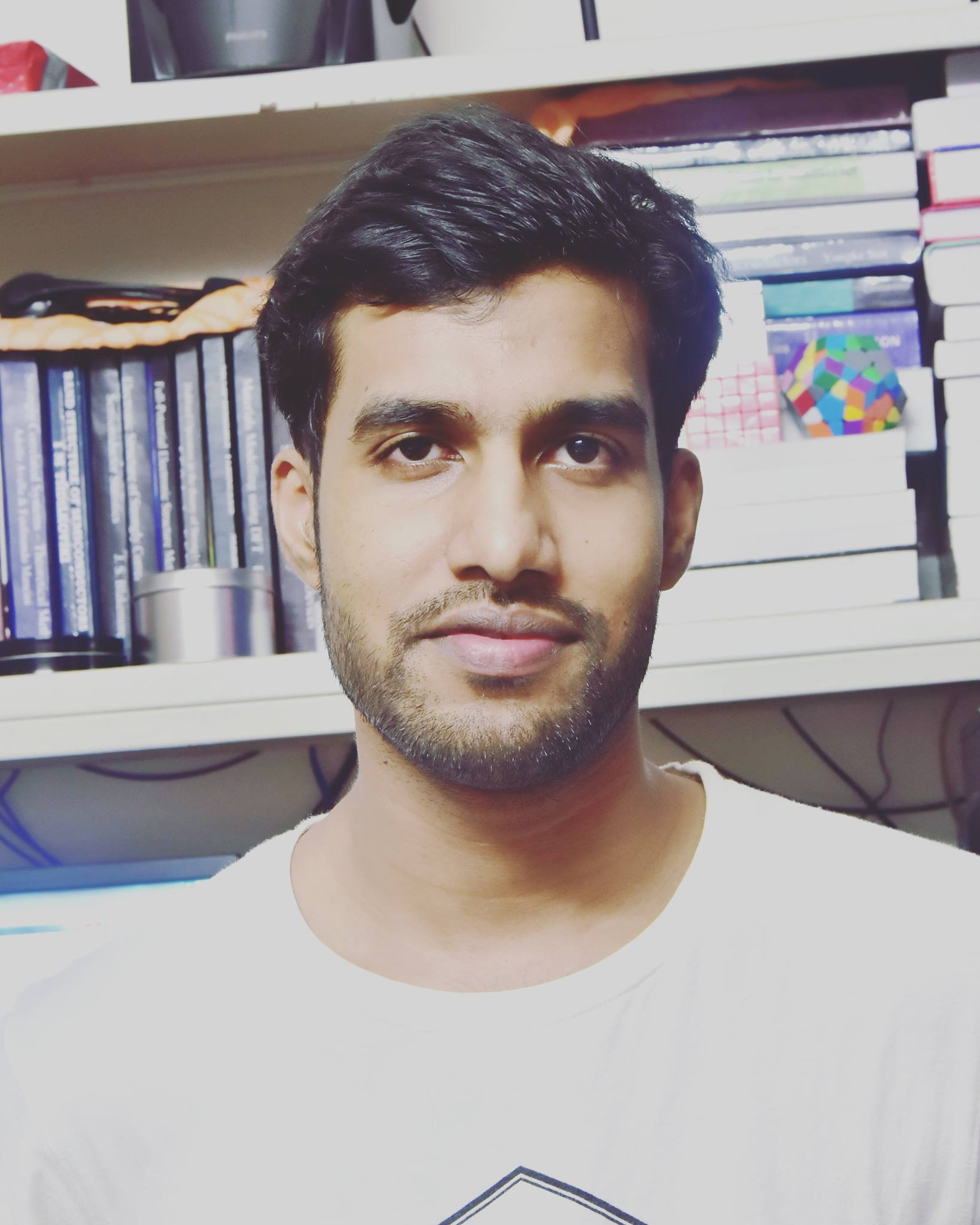}}]{Arijit Mondal} 
received the B.Tech. degree from National Institute of Technology, Durgapur, followed by M.E. and Ph.D from Indian Institute of Science. His research interests include quantum error correction code circuits and VLSI architectures for error correction codes. He is currently working as a Posdoctoral Associate at the University of Minnesota.
\end{IEEEbiography}

\begin{IEEEbiography}[{\includegraphics[width=1in,height=1.25in,clip,keepaspectratio]{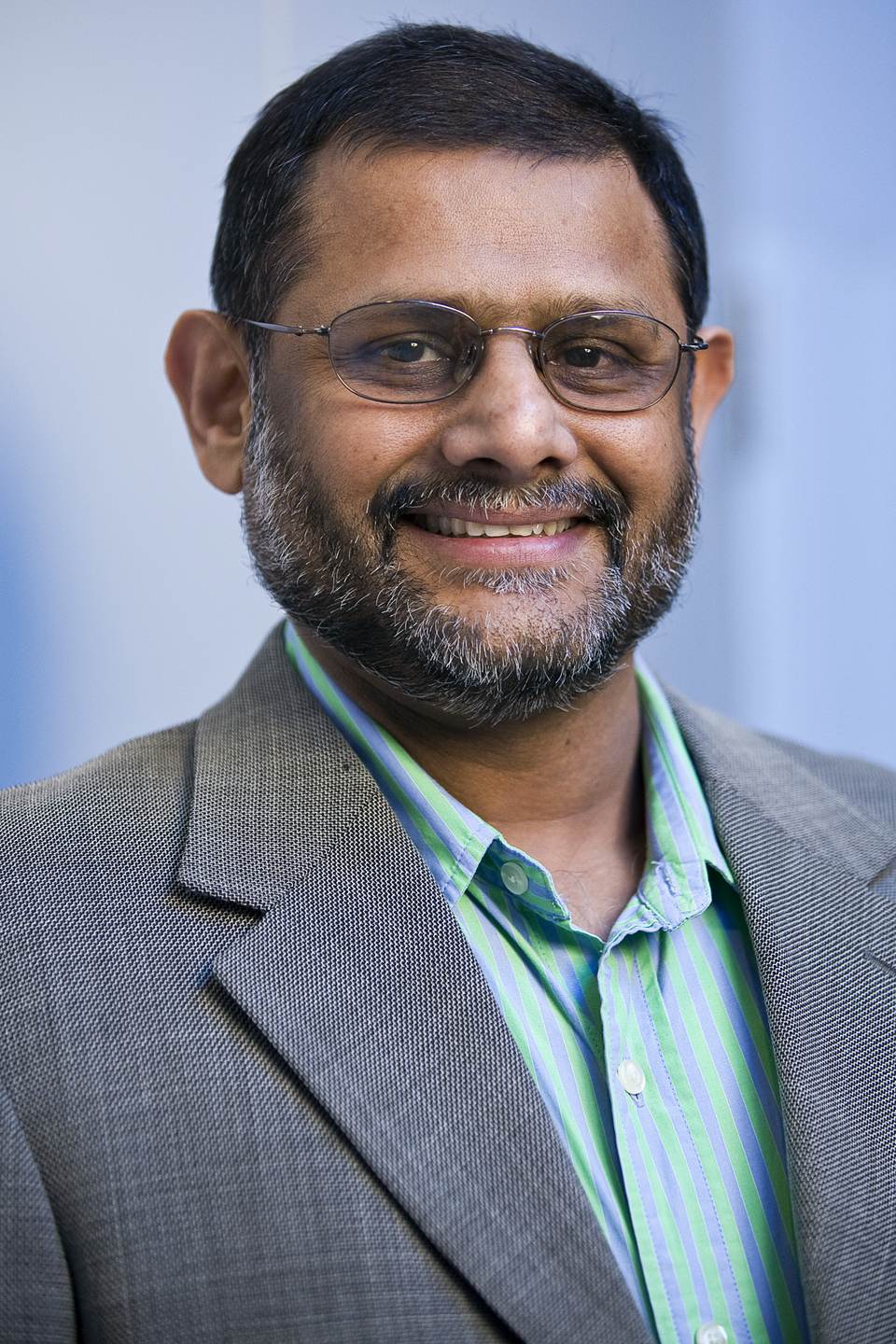}}]{Keshab K. Parhi} (Fellow, IEEE)
received the B.Tech. degree from the Indian Institute
of Technology (IIT), Kharagpur, in 1982, the M.S.E.E. degree from the
University of Pennsylvania, Philadelphia, in 1984, and the Ph.D.
degree from the University of California, Berkeley, in 1988. He has
been with the University of Minnesota, Minneapolis, since 1988, where
he is currently Erwin A. Kelen Chair and Distinguished McKnight University Professor
in the Department of
Electrical and Computer Engineering. He has published over 700 papers,
is the inventor of 35 patents, and has authored the textbook VLSI
Digital Signal Processing Systems (Wiley, 1999). His current research
addresses VLSI architecture design of machine learning and signal processing systems,
hardware security, and data-driven neuroengineering and neuroscience.
Dr. Parhi is the recipient of numerous awards including the
2017 Mac Van Valkenburg award and the 2012 Charles A. Desoer Technical
Achievement award from the IEEE Circuits and Systems Society, the 2003
IEEE Kiyo Tomiyasu Technical Field Award, and a Golden Jubilee medal
from the IEEE Circuits and Systems Society in 2000. He served as the
Editor-in-Chief of the IEEE Trans. Circuits and Systems, Part-I during
2004 and 2005. He is a
Fellow of the American Association for the Advancement of Science
(AAAS), the Association for Computing Machinery (ACM), the American Institute of
Medical and Biological Engineering (AIMBE), and the National Academy of Inventors (NAI).
\end{IEEEbiography}

\end{document}